\documentclass[conference]{IEEEtran}
\IEEEoverridecommandlockouts
\usepackage{cite}
\usepackage{amsmath,amssymb,amsfonts}
\usepackage{algorithmic}
\usepackage{graphicx}
\usepackage{textcomp}
\usepackage{xcolor}
\usepackage{subfig}
\usepackage{amsthm}
\theoremstyle{remark}
\newtheorem{remark}{Remark}

\def\BibTeX{{\rm B\kern-.05em{\sc i\kern-.025em b}\kern-.08em
    T\kern-.1667em\lower.7ex\hbox{E}\kern-.125emX}}
\begin{document}

\title{Deep Reinforcement Learning-Based Decision-Making Strategy Considering User Satisfaction Feedback in Demand Response Program\\
\thanks{This work was partly supported by Smart Grid-National Science and Technology Major Project 2024ZD0802100 and the National Natural Science Foundation of China under Grant 62373290 and Grant 62173256.}
}

\author{
\IEEEauthorblockN{Xin Li}
\IEEEauthorblockA{
  Department of Artificial Intelligence and Automation\\
  School of Electrical Engineering and Automation\\
  Wuhan University, P. R. China\\
  lixin9905@whu.edu.cn
}
\and
\IEEEauthorblockN{Li Ding*}
\IEEEauthorblockA{
  Department of Artificial Intelligence and Automation\\
  School of Electrical Engineering and Automation\\
  Wuhan University, P. R. China\\
  liding@whu.edu.cn
}
\and
\IEEEauthorblockN{Qiao Lin}
\IEEEauthorblockA{
  Department of Artificial Intelligence and Automation\\
  School of Electrical Engineering and Automation\\
  Wuhan University, P. R. China\\
  linqiao@whu.edu.cn
}
\and
\IEEEauthorblockN{Zhen-Wei Yu}
\IEEEauthorblockA{
  Department of Artificial Intelligence and Automation\\
  School of Electrical Engineering and Automation\\
  Wuhan University, P. R. China\\
 zhenweiyu@whu.edu.cn
}
}

\maketitle

\begin{abstract}
Demand response providers (DRPs) are intermediaries between the upper-level distribution system operator and the lower-level participants in demand response (DR) programs. Usually, DRPs act as leaders and determine electricity pricing strategies to maximize their economic revenue, while end-users adjust their power consumption following the pricing signals. However, this profit-seeking bi-level optimization model often neglects the satisfaction of end-users participating in DR programs. In addition, the detailed mathematical models underlying user decision-making strategy and satisfaction evaluation mechanism are typically unavailable to DRPs, posing significant challenges to conventional model-based solution methods. To address these issues, this paper designs a user-side satisfaction evaluation mechanism and proposes a multi-branch temporal fusion twin-delayed deep deterministic policy gradient (MBTF-TD3) reinforcement learning algorithm. User satisfaction feedback is incorporated into the reward function via a dynamically adjusted penalty term. The proposed MBTF structure effectively extracts temporal feature dependencies in the time-series observation data, and the dynamically adjusted penalty function successfully enhances the overall satisfaction level of users. Several experiments are conducted to validate the performance and the effectiveness of our proposed solution algorithm. 
\end{abstract}

\begin{IEEEkeywords}
Demand response, deep reinforcement learning, user satisfaction feedback, reward shaping.\end{IEEEkeywords}

\section{Introduction}
In recent years, the growing power shortages and the increasing penetration of renewable energy sources have highlighted the importance of demand response (DR) programs. These programs encourage consumers to adjust their power consumption behaviors in response to electricity pricing signals, thereby enhancing the stability and flexibility of smart grids. DR programs generally fall into two categories. The price-based DR adjusts users' power consumption by time-varying electricity price signals~\cite{RN76}, while the incentive-based DR provides additional rewards for users' power response~\cite{RN75}. 

Demand Response Providers (DRPs) play a critical role by connecting upper-level Distribution System Operators (DSOs) with lower-level end-users, fully exploiting the small-scale response resources~\cite{RN24}.  
Considering the natural characteristics of DRPs and end-users, interactions between these entities are typically modeled using a bi-level decision-making framework. Game-theoretic approaches, particularly Stackelberg games, are widely used to describe the hierarchical relationship, where DRPs are leaders setting pricing strategies, and end-users are followers adjusting their consumption accordingly~\cite{RN81, RN84}. Occasionally, a Nash game is applied among lower-level participants, resulting in a Stackelberg-Nash game model~\cite{RN80}. 
The bi-level optimization model treats the decision-making process of DRPs and end-users as a nested optimization problem~\cite{RN24, RN79,RN1090, RN999}. At the upper level, DRPs develop pricing strategies to maximize their benefits, while at the lower level, end-users optimize their electricity consumption behavior based on this price. The optimization framework clearly describes the optimal response relationships. This paper focuses on price-based DR programs and formulates a bi-level optimization problem to model the decision-making process between DRPs and end-users.

Most existing studies formulate profit-seeking bi-level models, where DRPs choose the economic optimal decision. However, during periods of low photovoltaic (PV) generation or high DSO electricity prices, profit-seeking DRPs may simply raise the selling price to obtain a relatively higher revenue. Although users give their “optimal" response to the price signal, their internal satisfaction with the program may reach a low level. Therefore, it is essential to incorporate feedback on user satisfaction into the DRPs' decision-making process. Some literature offers user satisfaction evaluation mechanisms. For instance, Ref.~\cite{WOS:000788363500006} uses the difference between the electricity consumption variations for the changed price to evaluate users' satisfaction. Ref.~\cite{WOS:000731149800020} designs a satisfaction function related to users' demand response participation ratio. Nevertheless, consumers typically have complex, multifaceted, and somewhat vague perceptions of product satisfaction. Single-dimensional or overly precise evaluation mechanisms are insufficient for simulating users' experiences with the programs.

In terms of solution methodologies, the existing approaches can be roughly divided into model-based and model-free categories. Model-based methods,  such as the Karush–Kuhn–Tucker transformation approach~\cite{RN136, RN137, RN138}, usually assume full knowledge of end-users' decision-making models. This assumption is impractical due to privacy concerns and the complex, diverse nature of user behaviors. Recent advancements in deep reinforcement learning (DRL) provide solutions to problems without explicitly modeling the environment. DRL agents learn optimal strategies through interaction with a black-box environment, effectively simulating the unknown dynamics of end-user responses. Numerous papers utilize DRL algorithms to solve the model-free optimization problem in DR programs~\cite{RN79, RN10, RN81, RN6, RN48,10602054}. For instance, Ref.~\cite{RN79} addressed the bi-level electric vehicle management problem using a safe DRL algorithm. Ref.~\cite{RN10} proposes a DRL-based dynamic pricing strategy to address the duck curve phenomenon through reward shaping. 

Motivated by these developments, this paper proposes a multi-branch temporal fusion twin-delayed
deep deterministic policy gradient (MBTF-TD3) algorithm, where the MBTF architecture uses Long Short-Term Memory (LSTM) networks effectively to capture the temporal dependencies in sequential data, such as PV generation and DSO electricity price. Simultaneously, the dynamically adjusted penalty function penalizes deviations from desired satisfaction levels, enabling the DRP to generate operational strategies in alignment with economic and user satisfaction considerations. The main contributions of this paper come from the following aspects: 
\begin{enumerate}
    \item This paper designs a user satisfaction evaluation mechanism that effectively assesses end-user satisfaction, which simulates consumers' multifaceted and vague perceptions of DR programs. 
    \item The user satisfaction feedback forms a dynamically adjusted penalty function integrated into the reward function, effectively balancing economic revenue and user satisfaction.
    \item A MBTF feature extractor is adopted to capture temporal dependencies in sequential observations, improving the DRP’s adaptability and performance in dynamic environments.
\end{enumerate}

\section{Problem Formulation}
This section formulates a bi-level optimization problem, representing the interaction between the DRP and end-users. Additionally, a mathematical model for evaluating users' satisfaction levels is introduced. 
\subsection{Response Model of End-Users}
\subsubsection{Decision Model}
Power users purchase electricity directly from the DRP. When the DRP establishes the electricity price $\lambda_t$, end-users change their power consumption according to their decision model. Typically, a welfare maximization problem models users' behavior~\cite{9937161}. 
\begin{equation}
\label{eq:user decision model}
\begin{aligned}
   f^i(\lambda_t) &=  \max_{d_t^i}   \; \sum_t{ { U_i(d_t^i)- \lambda_t d_t^i  }} \\
    \text{s.t.} \; &\underline{d_t^i} \leq d_t^i \leq \bar{d_t^i}, \forall{i,t} 
\end{aligned}
\end{equation}
where the utility function for user $i$, $U_i(d_t^i) = u_{a}^i(d_t^i)^2 + u_b^i d_t^i$, is concave in the feasible region with $u_a^i<0,u_b^i>0$~\cite{7063260}. The parameters $\bar{d_t^i}$ and $\underline{d_t^i}$ represent the upper and lower bounds of user $i$'s power consumption, respectively.
\subsubsection{Satisfaction Evaluation Mechanism}
Users' satisfaction regarding their response decisions relates to multiple aspects. Let $I_t^i$, $V_t^i$, and $L_t^i$ denote the deviation rate from users' ideal demand, the rate of power consumption change between consecutive periods, and the up/down limits satisfaction. The user $i$'s  satisfaction level at period $t$, denoted as $C_t^i$, is mathematically formulated as follows:
 \begin{equation}
\begin{aligned}
        c_t^i & = 10 -\omega_1 I_t^i - \omega_2 V_t^i - (10- \omega_1-\omega_2) L_t^i, \\
         C_t^i & = \lfloor c_t^i \rfloor, 
\end{aligned}
 \end{equation}
where
\begin{equation}
    \begin{aligned}
        I_t^i & = \left|d_t^i - {d_t^i}^{ideal}\right|/{\max \left(\bar{d_t^i} - {d_t^i}^{ideal},{d_t^i}^{ideal} - \underline{d_t^i} \right)}\\
        V_t^i &= \left|{d_t^i - d_{t-1}^i} \right|/  \left|\bar{d_t^i} - \underline{d_t^i} \right| \\
        L_t^i &= 
        \begin{cases} 
        1 & \text{if}\; \left|d_t^i-\bar{d_t^i}\right|\leq \epsilon \; \text{or}\; \left|d_t^i- \underline{d_t^i} \right| \leq \epsilon, \\
        0 & \text{otherwise}.
        \end{cases}
            \end{aligned}
        \end{equation}
It can be seen that the indices $I_t^i$, $V_t^i$, and $L_t^i$ are normalized to the interval $[0,1]$. The floor function $\lfloor c_t^i \rfloor$ is employed to reflect users' vague perceptions of satisfaction level. Thus, the satisfaction level satisfies $C_t^i \in \mathbb{Z}, 0 \leq C_t^i\leq 10$. The satisfaction level is assumed to be revealed only after the power demand has been fulfilled.

\subsection{Decision Model of the DRP}
The DRP purchases electricity from the DSO as a price-taker and sells it to end-users as a price-maker. Besides that, the DRP coordinates renewable energy resources and battery energy systems (BESs). Therefore, the DRP's optimal decision-making model is formulated as follows: 
\begin{align}
\max_{\lambda_t,p_t^b} & \sum_t {\sum_i \lambda_t d_t^i - p_t^{DSO}\lambda_t^{DSO} - \alpha_b (p_t^b)^2 - \rho^{res} p_t^{neg} } \label{eq:objective} \\
\text{s.t.} \quad & \forall t: \notag \\
& k_1 \lambda_t^{DSO} \leq \lambda_t \leq k_2 \lambda_t^{DSO} \label{eq:price_bound} \\
& \lambda_{t-1} - \Delta \lambda_t \leq \lambda_t \leq \lambda_{t-1} + \Delta \lambda_t \label{eq:price_variation} \\
& p^b_{min} \leq p_t^b \leq p^b_{max} \label{eq:power_bound} \\
& Soc_t^{min} \leq Soc_t \leq Soc_t^{max} \label{eq:soc_bound} \\
& Soc_t = Soc_{t-1} + \frac{\max(0,p_t^b)\eta_{ch}}{Capacity} + \frac{\min(0,p_t^b)}{\eta_{dis} Capacity} \label{eq:soc_update} \\
& p_t^{res} - p^b_t - \sum_i d_t^i = p_t, \quad \label{eq:power_balance} \\
& p_t^{DSO} = \max(0,p_t) \label{eq:positive_power} \\
& p_t^{neg} = \max(0,-p_t) \label{eq:negative_power} \\
& d_t^i = \arg\max f^i(\lambda_t), \; \forall i \label{eq:user_response} \\
& C_t^{ave} \geq C^{b} \label{eq:satisfaction_constraint}
\end{align}

Eq.(\ref{eq:objective}) specifies the DRP's optimization objective, including revenue from electricity sales to end-users, electricity procurement costs from the DSO, the BES utilization costs, and penalties for not fully using renewable energy. Constraints (\ref{eq:price_bound})-(\ref{eq:price_variation}) represent the upper and lower bounds, as well as variation limits, in pricing decisions. Constraints (\ref{eq:power_bound})-(\ref{eq:soc_update}) specify the operational limits and state-of-charge ($Soc$) dynamics of the BES. Constraints (\ref{eq:power_balance})-(\ref{eq:negative_power}) enforce the balance between power generation and demand, where $p_t^{neg}$ denotes unused renewable energy and $\rho^{res}$ represents the associated penalty cost. Constraint (\ref{eq:user_response}) represents end-users' optimal responses to the pricing decisions at the lower-level optimization. Lastly, constraint (\ref{eq:satisfaction_constraint}) ensures a minimum average satisfaction level among users, where $C_t^{ave}$ is the average satisfaction at period~$t$ and $C^{b}$ is a predefined satisfaction threshold.

\section{Solution Methodology}
In this section, the bi-level problem is reformulated into a Markov Decision Process (MDP), and a dynamically adjusted penalty function for user satisfaction is proposed.

\subsection{MDP formulation}
The MDP is defined as a tuple $<\mathcal{S},\mathcal{A},P,r>$, where $\mathcal{S}$ is the state space, $\mathcal{A}$ is the action space, $P(s'|s,a):\mathcal{S}\times \mathcal{A}\times \mathcal{S}:\xrightarrow{} [0,1]$ denotes the state transition probability, and $r(s,a): \mathcal{S}\times \mathcal{A} \xrightarrow{}\mathbb{R} $ is the reward function. The policy function $\pi(a,s):\mathcal{S}\times \mathcal{A}\xrightarrow{}[0,1]$ determines the probability of choosing action $a \in \mathcal{A} $ in state $s\in \mathcal{S}$, satisfying $\sum_{a\in \pi} \pi(s,a) = 1$. A detailed description of the MDP is given below.

\subsubsection{States} The state includes historical data from the past $T_{his}$ periods and the predicted data in the future $T_{pre}$ periods on PV generation and DSO electricity price, the average user satisfaction level from the previous period, the battery state-of-charge $Soc$ from the previous period, and the time-awareness state for the current period. The state space for $\forall t \in T$ is defined as 
\begin{equation}
    s_t = \left [   ({p}^{res}_{k} )_{k=t-T_{his}+1}^{t+ T_{pre}},{({\lambda}_{k}^{DSO})_{k=t-T_{his}+1}^{t+ T_{pre}}},c_{t-1}^{ave},Soc_{t-1},\bf{ta_{t}} \right ],
\end{equation}
where ${\bf{ta_t}}$ represents the vector of sine-cosine encoded time information, enhancing the learning of potential temporal relationships. Specifically, the sine/cosine encoding of hour $h_t$, week $w_t$, and month $m_t$ are defined as~\cite{sage2025deep}
\begin{equation*}
\begin{aligned}
        h_{t,sin}  & = \sin(2\pi \cdot \frac{h_t}{23}),  h_{t,cos} = \cos(2\pi \cdot \frac{h_t}{23}), \\
        w_{t,sin} &= \sin(2\pi \cdot \frac{w_t}{51}),  w_{t,cos} = \cos(2\pi \cdot \frac{w_t}{51}), \\
        m_{t,sin} &= \sin(2\pi \cdot \frac{m_t}{11}),  m_{t,cos} = \cos(2\pi \cdot \frac{m_t}{11}).
\end{aligned}
\end{equation*}
Then, ${\bf{ta_t}} = [h_{t,sin},h_{t,cos},w_{t,sin},w_{t,cos},m_{t,sin},m_{t,cos}]$.

\subsubsection{Actions} Upon observing state $s_t$, the DRP decides the electricity selling price and the charge/discharge power. Thus, the action $a_t$ is represented by 
\begin{equation*}
    a_t = \left[ \lambda_t,p_t^b \right] ,  \forall t \in T.
\end{equation*}

\subsubsection{Rewards} After establishing the electricity price, the DRP receives load and satisfaction feedback from end-users. Then, the DRP computes the sell-electricity revenue, the BES utilization cost, the buy-electricity cost, or the wasted renewable energy penalty, and finally computes the reward by
\begin{equation}
\begin{aligned}
        r_t = & {\sum_i {\lambda_t d_t^i}  - p_t^{DSO}\lambda_t^{DSO} - \alpha_b (p_t^b)^2 - \rho^{res} p_t^{neg}} \\
        & - Pen_t(C_t^i), \forall i \in T,
\end{aligned}
\end{equation}
where the penalty function $Pen(C_t^i)$ associated with the user satisfaction level will be designed in the next subsection.

\subsubsection{Transitions} The transition probability $P(s'|s, a)$ remains implicit, as the system dynamics, including the decision model of end-users and uncertainties from the DSO price and PV generation, are unknown to the DRP. Consequently, a model-free DRL algorithm is applied to tackle the proposed problem. 

\subsection{Dynamically Adjusted Penalty Function}
The DRP agent should be penalized if users' average satisfaction levels fall below a threshold. The average satisfaction level of end-users at period $t$ is calculated by 
\begin{equation}
    C_t^{ave} = \frac{1}{N_d*t} \sum_{k=1}^t \sum_{i=1}^{N_d} {C_k^i} ,\; \forall t \in T,
\end{equation}
Thus, Eq.~\eqref{eq:soc_bound} ensures that at $\forall t \in T$, users' average satisfaction levels are consistently lower-bounded by $C^b$.

Traditional penalty functions, including linear and squared forms, are respectively defined by:
\begin{equation}
    \begin{aligned}
        Pen_t & = \beta^{lin} \cdot \max(0,C^{b} -  C_t^{ave}), \\
        Pen_t& = \beta^{sqr} (C_t^{b} -  C_t^{ave})^2, 
    \end{aligned}
\end{equation}
where $\beta^{lin}$ is a constant penalty coefficient for users' insufficient satisfaction, and $\beta^{sqr}$ is a constant coefficient for any satisfaction deviations.
Inspired by the Lagrangian Relaxation method, a dynamic combined penalty function is proposed.
\begin{equation}
\begin{aligned}
      Pen_t &  = \frac{1}{2} \beta_t^{lin} \max(0, C_t^{b} -  C_t^{ave}) + \frac{1}{2} \beta_t^{sqr} (C_t^{b} -  c_t^{ave})^2 \\
      \beta_t^{lin} & = \max(0,\beta_t^{lin} + \eta^{lin}(C_t^{b} -  C_t^{ave})) \\
      \beta_t^{sqr}& = \beta_t^{sqr} + \eta^{sqr} |C_t^{b} -  C_t^{ave}|
\end{aligned}
\end{equation}
where $\beta_t^{lin}$ and $\beta_t^{sqr}$ dynamically adjust according to the satisfaction difference, and $\eta^{lin}$ and $\eta^{sqr}$ are the adjustment steps for $\beta_t^{lin}$ and $\beta_t^{sqr}$, respectively. Specifically, the dynamic squared penalty term rapidly intensifies the penalty strength in cases of large deviations from user satisfaction bounds, efficiently compelling the DRP to reduce deviations promptly. Meanwhile, the dynamic linear penalty term provides stable and continuous incentives in cases of small deviations.

\subsection{DRL Solution Algorithm}
With the MDP model formulated before, the DRP is set as a DRL agent, while the DSO electricity price, PV generation, end-users, and the BES compose the environment. To effectively handle continuous action spaces and the inherent complexity of this environment, the model-free DRL algorithm, TD3, is adopted as the decision-making strategy for the DRP agent. As an improvement of the classic deep deterministic policy gradient algorithm, TD3 stabilizes training by mitigating value overestimation through dual Q-networks and delayed policy updates. 

\begin{figure}
    \centering
    \includegraphics[width=0.7\linewidth]{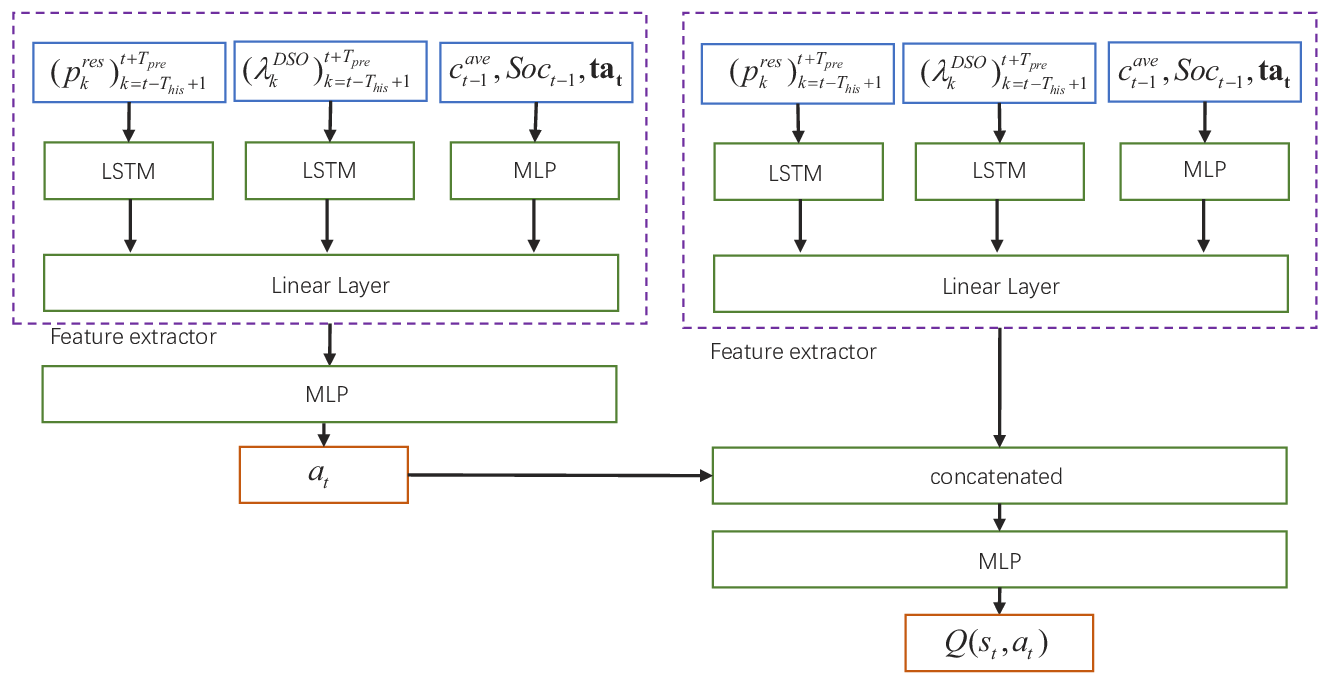}
    \caption{The neural network structure.}
    \label{fig:MBTF}
\end{figure}

Conventionally, TD3 policy networks rely on a multilayer perceptron (MLP) structure,  directly mapping states to corresponding actions. Given the extensive state space and distinct temporal characteristics of PV generation and DSO price data, our approach employs a multi-branch framework to handle different data types effectively. The detailed feature extractor structure is illustrated in Fig.~\ref{fig:MBTF}.  

The observations fed into the neural network consist of two types of data. First, the time-series data of PV generation $ ({p}^{res}_{k} )_{k=t-T_{his}+1}^{t+ T_{pre}}$ and DSO electricity price $({\lambda}_{k}^{DSO})_{k=t-T_{his}+1}^{t+ T_{pre}}$, which include both historical values and forecast values. These two time-series datasets are fed into two separate LSTM network branches to extract time-sequential features, each of them outputs a feature vector. Second, the non-sequential observation data ${[c_{t-1}^{ave}, Soc_t, \bf{ta_{t}}]}$, including average user satisfaction, the state-of-charge, and time-encoded vector, are processed through a conventional MLP branch. 
The extracted feature vectors from these three branches are then concatenated into a unified representation and passed through an additional linear layer to produce a compact feature and an integrated feature embedding. This representation is then forwarded into fully connected layers of the TD3 policy network, which ultimately determines the action~$a_t$. Additionally, the critic network, which estimates the value function $Q(s_t,a_t)$, utilizes the same structured feature extraction process. Specifically, it concatenates the extracted features with the action vector and processes them through an MLP branch to provide accurate value estimations.


\begin{remark}
The framework captures user heterogeneity by assigning individualized utility parameters and demand bounds, reflecting diverse consumption preferences and response behaviors.
\end{remark}
\begin{remark}
The dynamically adjusted penalty function is inspired by Lagrangian relaxation~\cite{fisher1981lagrangian}, enabling soft constraint enforcement and balancing economic performance with user satisfaction.
\end{remark}
\begin{remark}
In real-world deployment, model-free RL eliminates the need for explicit user models but still requires consumption data collection, which may face privacy regulations. Additionally, well-trained policies should maintain robustness under dynamic and uncertain power system conditions to ensure reliable real-time decision-making.
\end{remark}

\section{Case Studies}

\subsection{Experimental Settings}
The experiments are conducted in Python $3.10$, with the OpenAI Gym framework used to construct the self-defined environment and the stable-baseline3 Library used to implement DRL algorithms. 

\begin{figure}
    \centering
    \subfloat[]{
    \includegraphics[width=0.32\linewidth]{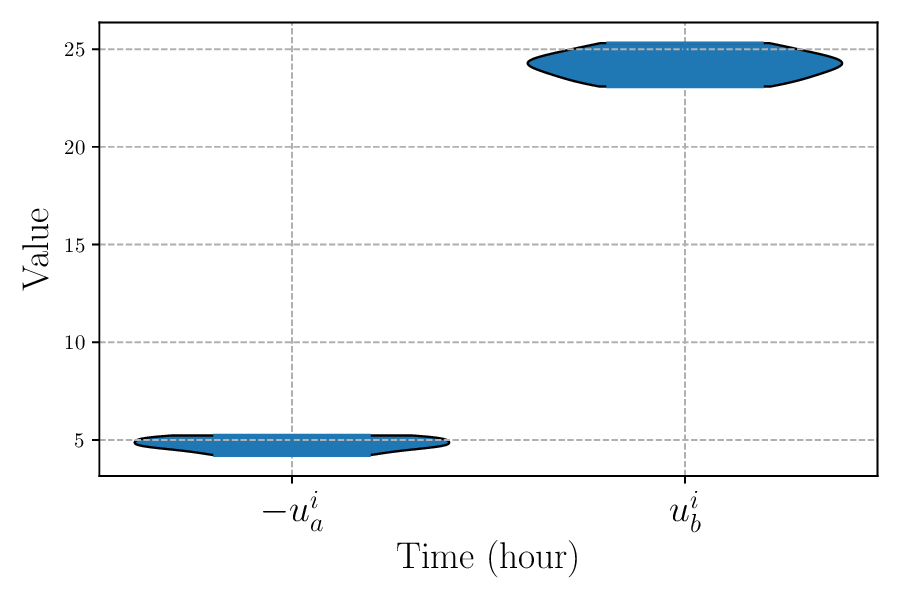}
    } 
        \subfloat[]{
    \includegraphics[width=0.32\linewidth]{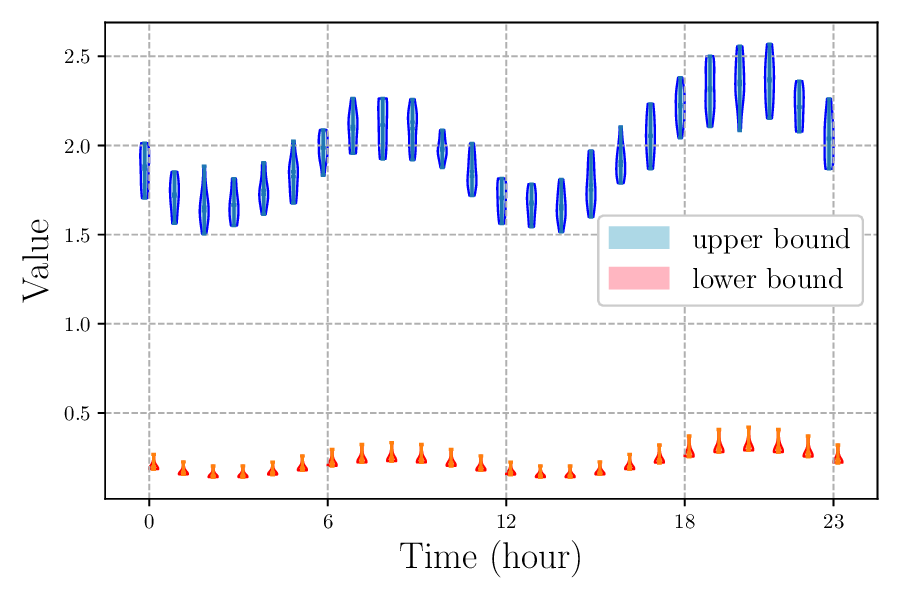}
    } 
    \caption{Parameter settings of end-users. (a) Utility function. (b) Power demand limits.}
    \label{fig:user-para}
\end{figure}

The PV generation data are sourced from Renewables.ninja~\cite{RenewablesNinja, STAFFELL20161224}, and the DSO price data are obtained from PJM hourly Real-time~\cite{PJMDataMiner}. These data are scaled appropriately to match the demand response program described in this study. System configurations for end-users are illustrated in Fig.~\ref{fig:user-para}. The parameter settings for DRL training and feature extractors are listed in Table~\ref{tab:mbtf_td3_params}. 

\begin{figure}
    \centering
    \includegraphics[width=0.6 \linewidth]{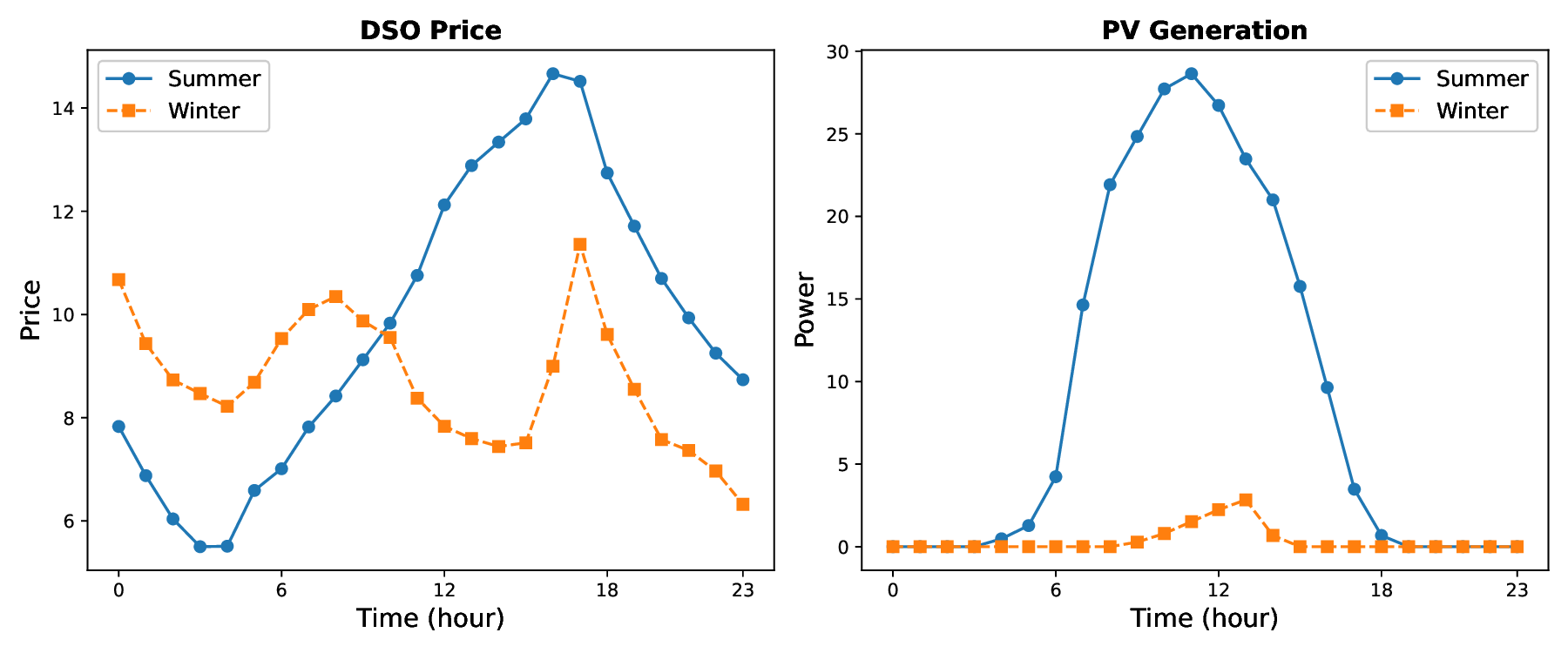}
    \caption{The DSO price and PV generation curves in summer/winter day.}
    \label{fig:prediction}
\end{figure}

To validate the effectiveness of the proposed algorithm, two representative days, one typical summer day (Oct.~$5^{th}$) and one typical winter day (Jan.~$15^{th}$), are selected. The associated DSO price and PV generation profiles for these days are depicted in Fig.~\ref{fig:prediction}. 

\begin{table}[t!]
\centering
\caption{Parameter Settings}
\label{tab:mbtf_td3_params}
\resizebox{0.8\columnwidth}{!}{
\begin{tabular}{lcl}
\hline
\textbf{Parameter} & \textbf{Value} & \textbf{Description} \\ \hline
Discount Factor ($\gamma$) & $0.99$ & Discount rate for future rewards \\ 
Learning Rate Actor & $10^{-5}$ & Step size for actor network updates \\ 
Learning Rate Critic & $10^{-5}$ & Step size for critic network updates \\ 
Batch Size & 100 & Number of samples per training batch \\ 
Replay Buffer Size & $10^6$ & Capacity of experience replay buffer \\ 
Policy Update Frequency & $2$ & Actor network updates every N critic updates \\ 
Target Update Rate ($\tau$) & $0.005$ & Soft update rate for target networks \\ 
\hline
\textbf{Feature Extractor Parameters} &  &  \\ \hline
PV Sequence Length & $32$ & Historical and prediction sequence length for PV data \\ 
DSO Sequence Length & $32$ & Historical and prediction sequence length for DSO data \\ 
PV LSTM Hidden Size & $16$ & Hidden state dimension for PV LSTM network \\ 
DSO LSTM Hidden Size & $16$ & Hidden state dimension for DSO LSTM network \\ 
MLP Input Dimension & $8$ & Dimension for scalar inputs (user satisfaction, $Soc$, time encoding) \\ 
MLP Hidden Dimension & $8$ & Hidden layer size for scalar information MLP \\ 
Combined Feature Dimension & $64$ & Output dimension after feature concatenation and linear layer \\ \hline
\end{tabular}
}
\end{table}


\subsection{Main Results}
\subsubsection{Algorithm Training Performance}
\begin{figure}
    \centering
    \subfloat[]{
    \includegraphics[width=0.6 \linewidth]{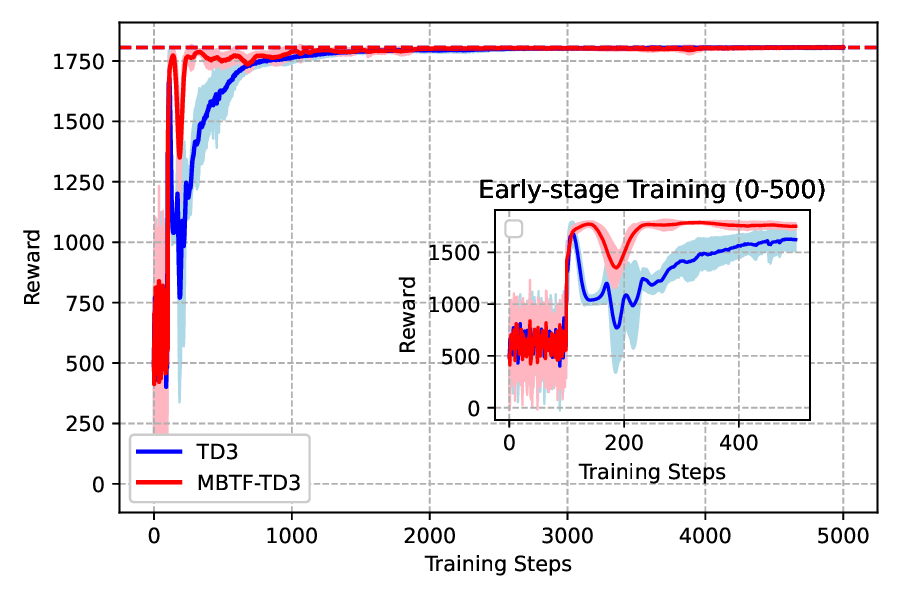}
    } \\
        \subfloat[]{
    \includegraphics[width=0.6 \linewidth]{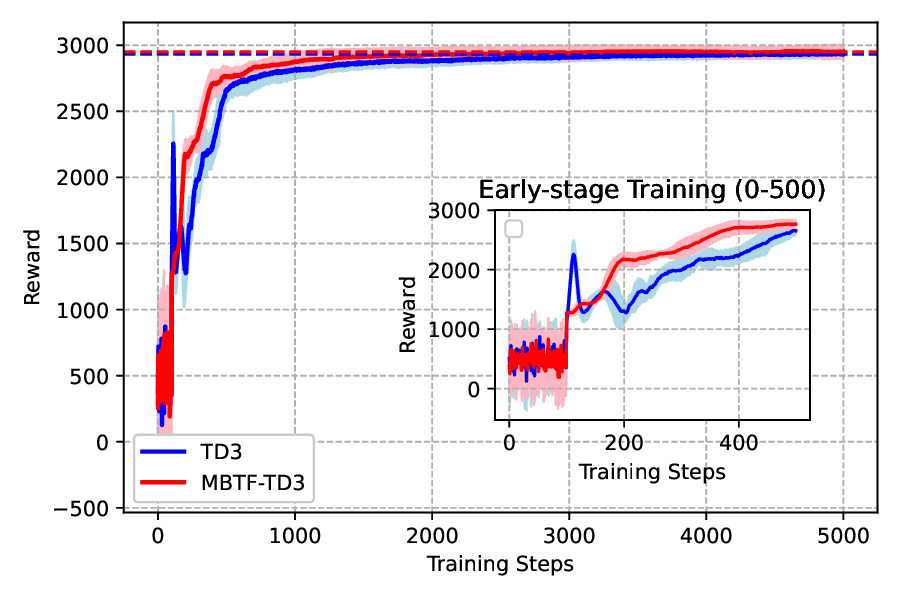}
    } \\
    \caption{Algorithm comparison results. (a) Winter day. (b) Summer day.}
    \label{fig:convergence}
\end{figure}

This part evaluates the performance of the proposed MBTF-TD3 algorithm compared to the classical TD3 algorithm. The training curves of these DRL algorithms are shown in Fig.~\ref{fig:convergence}. To ensure a fair comparison, we conduct $10$ independent training runs under identical initial environment conditions and training parameters, with each run using a different random seed. 
The convergence return value for the TD3 algorithm and MBFT-TD3 algorithm are $1805.85$ and $1806.32$ on the winter day and $2931.24$ and $2952.59$ on the summer day. In addition, the inset magnifies the early training phase to highlight differences in convergence speed. Results indicate that the proposed MBTF-TD3 algorithm exhibits better performance in convergence speed and achieves a higher return compared to the baseline TD3 algorithm.

\subsubsection{Users' Satisfaction Level Feedback Impact}
\begin{figure}
    \centering
  \includegraphics[width=0.7\linewidth]{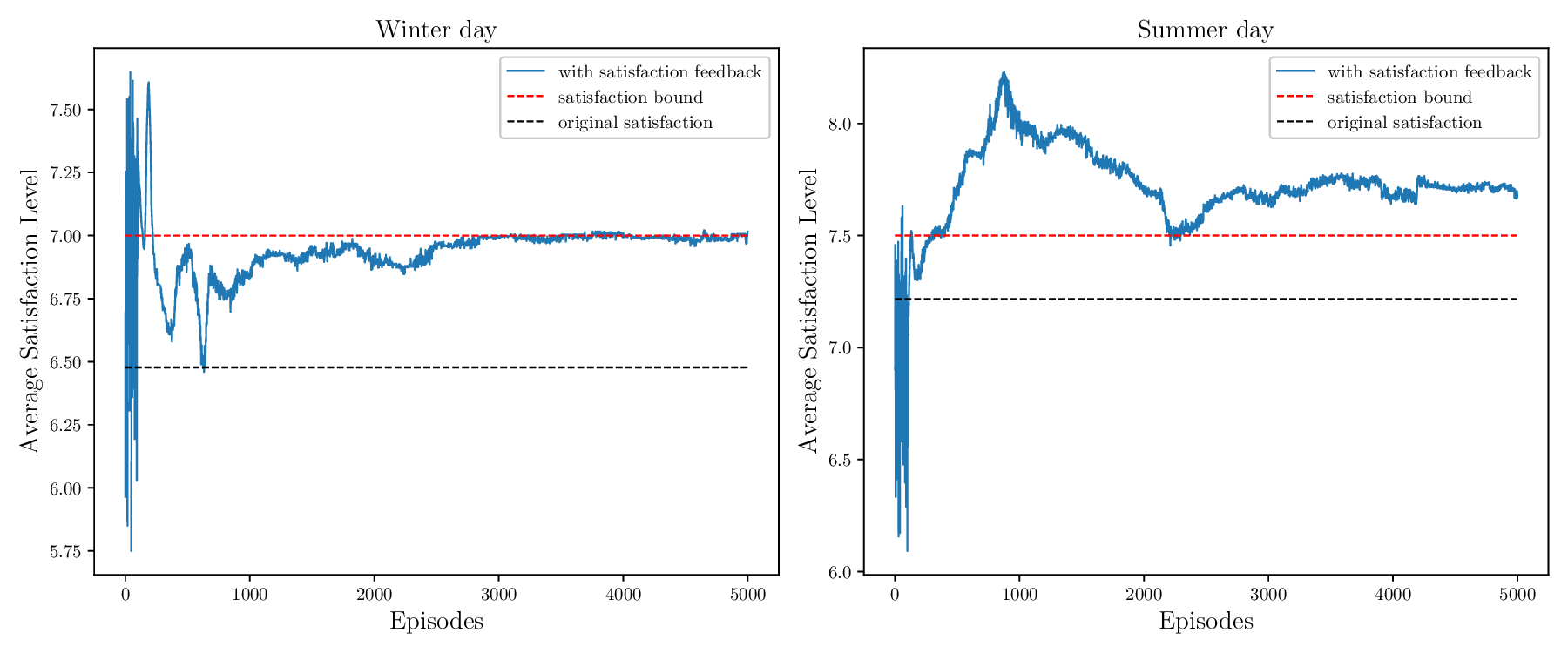}
    \caption{Average satisfaction levels during training phases.}
    \label{fig:penalty_term}
\end{figure}

\begin{figure}
    \centering
    \subfloat[]{
    \includegraphics[width=0.6\linewidth]{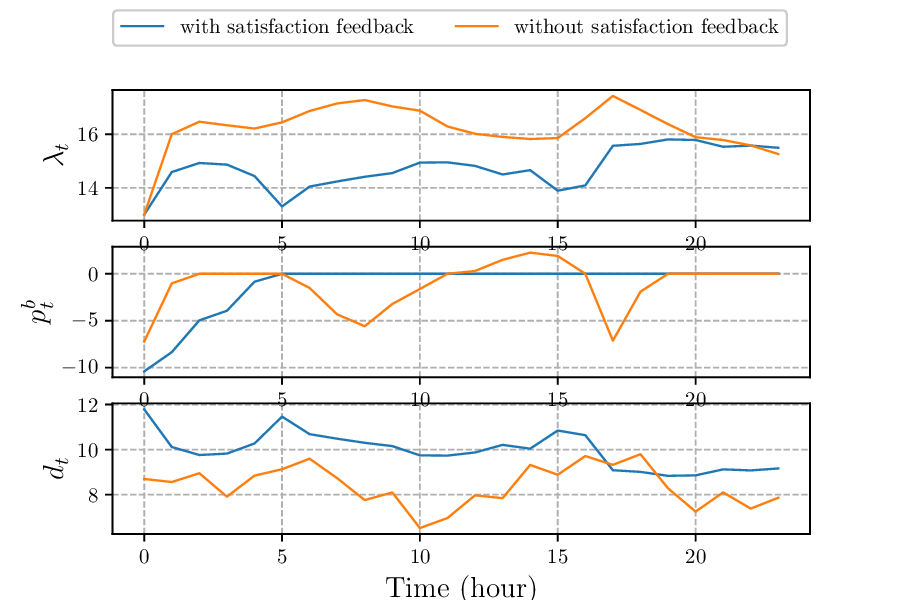}
    } \\
    \subfloat[]{
    \includegraphics[width=0.6 \linewidth]{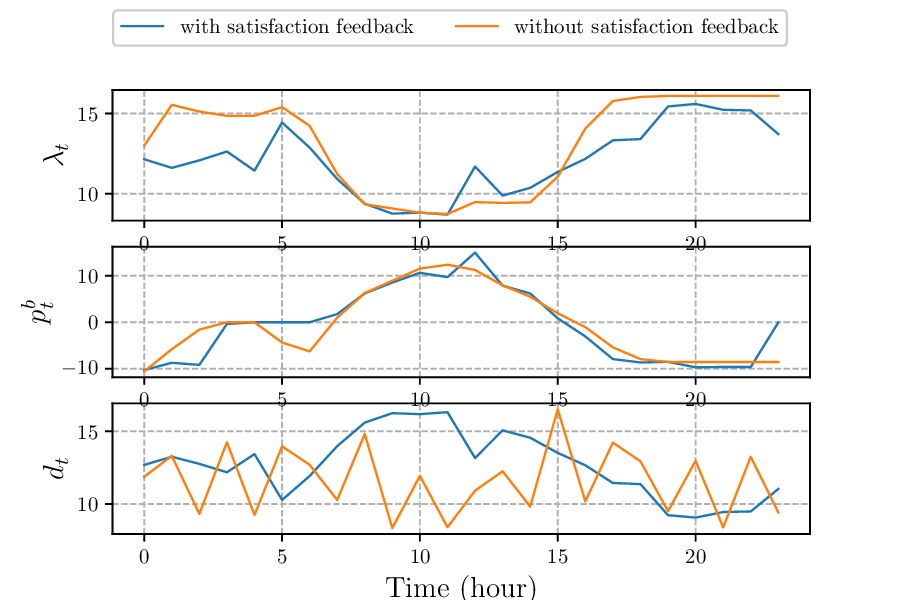}
    }
    \caption{Decision strategy comparison results. (a) Winter day. (b) Summer day.}
    \label{fig:decision-trategy}
\end{figure}

\begin{figure}
    \centering
     \subfloat[]{
    \includegraphics[width=0.35\linewidth]{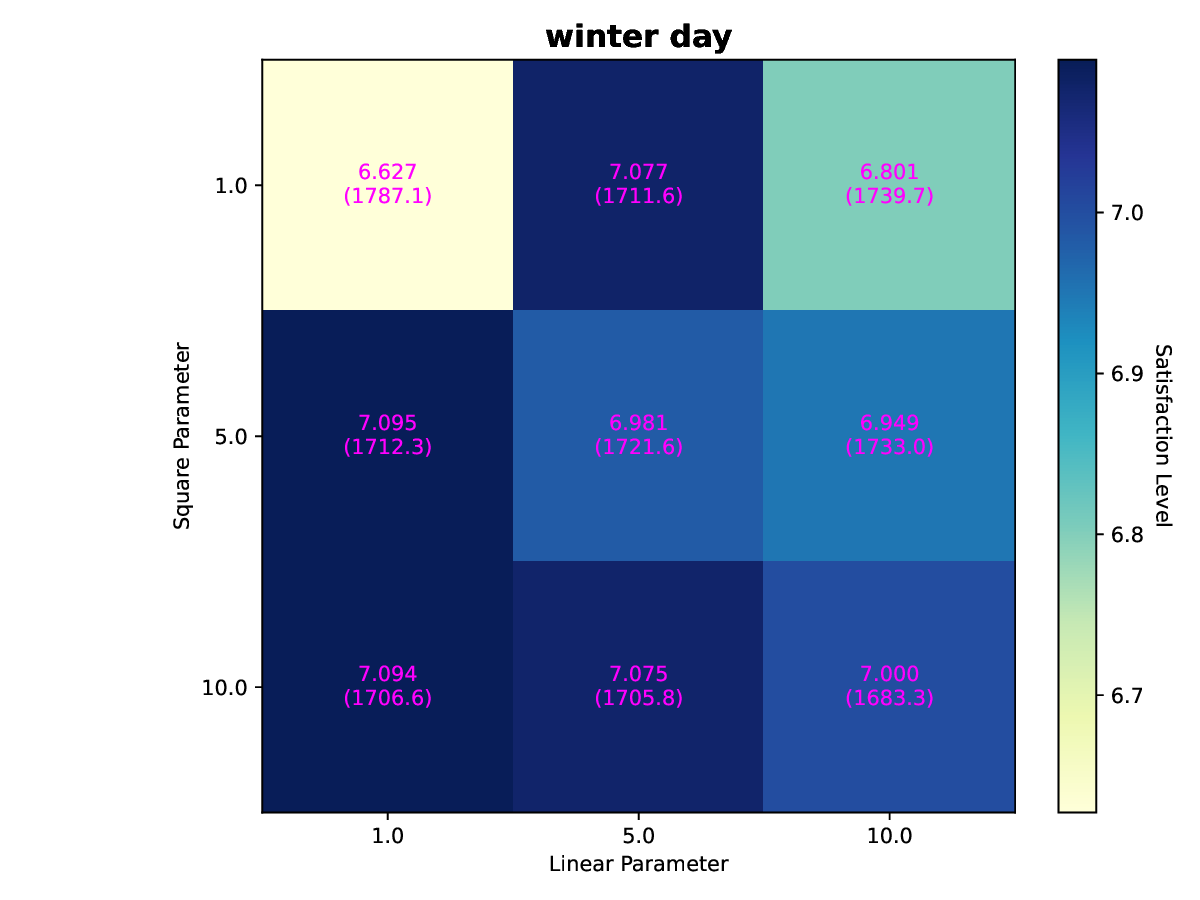}
    } 
    \subfloat[]{
    \includegraphics[width=0.35 \linewidth]{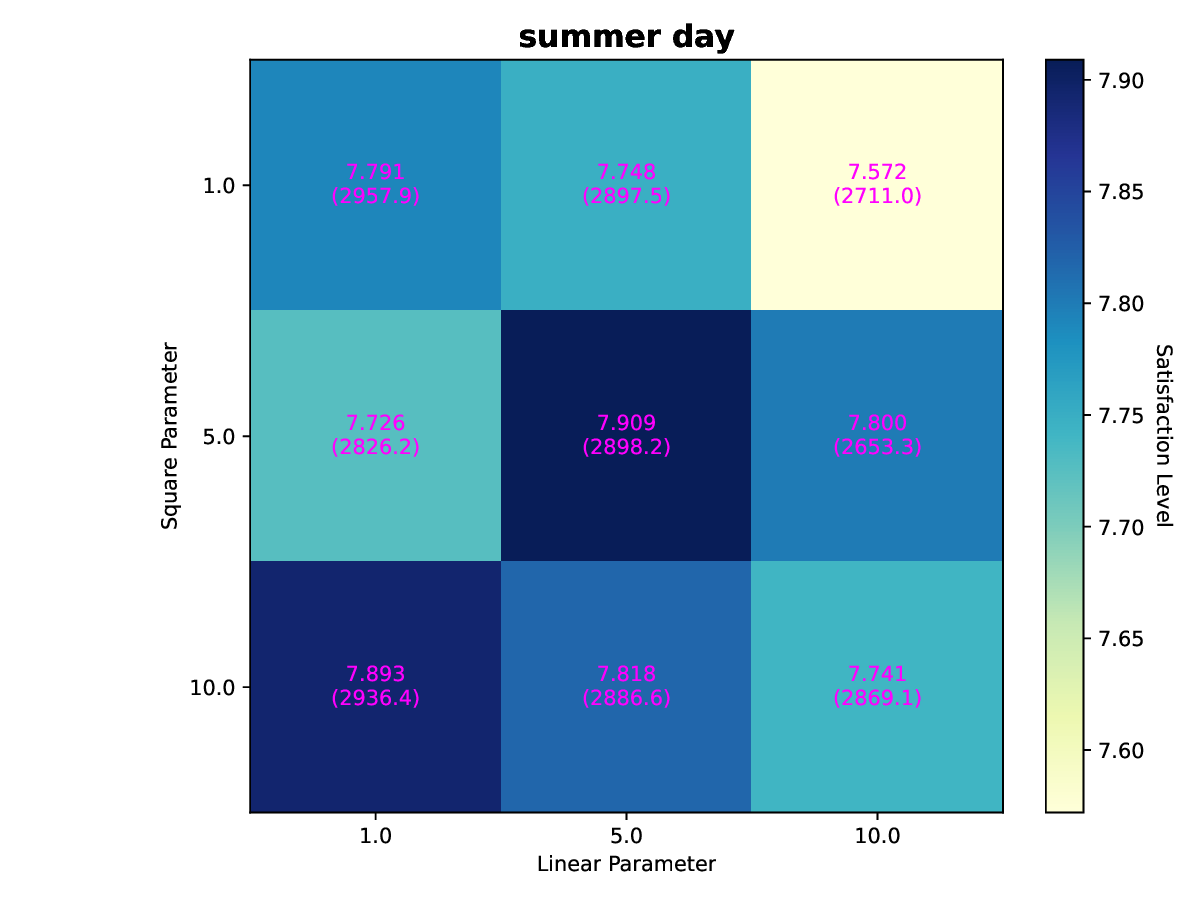}
    }
    \caption{Parameter sensitivity of penalty coefficient adjust step.}
    \label{fig:parameter-sensitivity}
\end{figure}

Users' day-average satisfaction levels during the training process are shown in~Fig.~\ref{fig:penalty_term}. The penalty parameters are $\eta^{lin}=5.0, \eta^{sqr}=1.0$ for winter day, $\eta^{lin}=5.0, \eta^{sqr}=5.0$ for summer day, and $\beta^{lin}=10,\beta^{sqr} = 20$ for the two days. It can be seen that the converged day-average satisfaction level is close and even exceeds the bound. The impact of incorporating user satisfaction feedback into the decision-making process is presented in Fig.~\ref{fig:decision-trategy}. When user satisfaction feedback is integrated, the DRP generates a relatively lower electricity pricing signal, and the BES tends to discharge to provide power, particularly during winter days, effectively promoting users' satisfaction levels. Next, we discuss the parameter sensitivity of the penalty coefficient adjust step $\eta^{lin}$ and $\eta^{sqr}$, with results shown in Fig.~\ref{fig:parameter-sensitivity}. On winter days, due to insufficient PV generation, increasing the satisfaction level will reduce the economic revenue. Thus, the average satisfaction level converges to the bound. On summer days, the average satisfaction level may exceed the bound to achieve higher revenue.

\subsubsection{Comparison}
Ref.~\cite{WOS:000788363500006,WOS:000731149800020} considered users' satisfaction, but the evaluation mechanism is single-dimensional, while our proposed mechanism combines the demand deviation, ideal demand, and limits satisfaction. Ref.~\cite{RN24} investigated a similar problem from the perspective of data efficiency and proposed an off-line DRL algorithm. Differing from that, our work focuses on the impact of user feedback on DRP decision-making and introduces an online DRL framework with satisfaction-aware reward shaping.

\section{Conclusion}
This paper addresses the users' satisfaction feedback issue in the DR program between DRPs and multiple end-users. Traditional profit-seeking bi-level optimization models inadequately capture users' satisfaction within the response program. Additionally, the model-based solution methods usually require detailed knowledge of users' decision models. To bridge this gap, we design a user-side satisfaction level evaluation mechanism to reflect users' overall satisfaction and a model-free MBTF-TD3 algorithm for the DRP to generate optimal pricing and charging/discharging strategies. Users' satisfaction feedback is integrated into the DRP's decision-making process through a dynamically adjusted penalty function in the reward function. The proposed MBTF architecture effectively captures the temporal dependencies in the observation data, thereby enhancing the DRP’s decision-making performance. Experimental results validate that the proposed method improves user satisfaction while maintaining economic performance.

\bibliographystyle{IEEEtran}
\bibliography{ICPST.bib}

\vspace{12pt}

\end{document}